\documentclass[final]{cvpr}

\usepackage{times}
\usepackage{epsfig}
\usepackage{graphicx}
\usepackage{amsmath}
\usepackage{amssymb}
\usepackage[table,x11names]{xcolor}

\usepackage[pagebackref=true,breaklinks=true,colorlinks,bookmarks=false]{hyperref}

\def\cvprPaperID{****} 
\def\confYear{CVPR 2021}
\definecolor{grayhighlight}{RGB}{213,229,255}

\newsavebox\CBox
\def\textBF#1{\sbox\CBox{#1}\resizebox{\wd\CBox}{\ht\CBox}{\textbf{#1}}}

\begin{document}

\title{\vspace{-4mm}
Learned Smartphone ISP on Mobile NPUs with Deep Learning,\\ Mobile AI 2021 Challenge: Report}

\author{
Andrey Ignatov \and Cheng-Ming Chiang \and Hsien-Kai Kuo \and Anastasia Sycheva \and Radu Timofte \and
Min-Hung Chen \and Man-Yu Lee \and Yu-Syuan Xu \and Yu Tseng \and
Shusong Xu \and Jin Guo \and
Chao-Hung Chen \and Ming-Chun Hsyu \and Wen-Chia Tsai \and Chao-Wei Chen \and
Grigory Malivenko \and
Minsu Kwon \and Myungje Lee \and Jaeyoon Yoo \and Changbeom Kang \and Shinjo Wang \and
Zheng Shaolong \and Hao Dejun \and Xie Fen \and Feng Zhuang \and
Yipeng Ma \and Jingyang Peng \and Tao Wang \and Fenglong Song \and
Chih-Chung Hsu \and Kwan-Lin Chen \and Mei-Hsuang Wu \and
Vishal Chudasama \and Kalpesh Prajapati \and Heena Patel \and Anjali Sarvaiya \and Kishor Upla \and Kiran Raja \and Raghavendra Ramachandra \and Christoph Busch \and
Etienne de Stoutz}
\maketitle

\begin{abstract}
As the quality of mobile cameras starts to play a crucial role in modern smartphones, more and more attention is now being paid to ISP algorithms used to improve various perceptual aspects of mobile photos. In this Mobile AI challenge, the target was to develop an end-to-end deep learning-based image signal processing (ISP) pipeline that can replace classical hand-crafted ISPs and achieve nearly real-time performance on smartphone NPUs. For this, the participants were provided with a novel learned ISP dataset consisting of RAW-RGB image pairs captured with the Sony IMX586 Quad Bayer mobile sensor and a professional 102-megapixel medium format camera. The runtime of all models was evaluated on the MediaTek Dimensity 1000+ platform with a dedicated AI processing unit capable of accelerating both floating-point and quantized neural networks. The proposed solutions are fully compatible with the above NPU and are capable of processing Full HD photos under 60-100 milliseconds while achieving high fidelity results. A detailed description of all models developed in this challenge is provided in this paper.

\end{abstract}
{\let\thefootnote\relax\footnotetext{%
\hspace{-5mm}$^*$Andrey Ignatov, Cheng-Ming Chiang, Hsien-Kai Kuo and Radu Timofte are the main MAI 2021 challenge organizers \textit{(andrey@vision.ee.ethz.ch, jimmy.chiang@mediatek.com, hsienkai.kuo@mediatek.com, radu.timofte @vision.ee.ethz.ch)}. The other authors participated in the challenge. \\ Appendix \ref{sec:apd:team} contains the authors' team names and affiliations. \vspace{2mm} \\ Mobile AI 2021 Workshop website: \\ \url{https://ai-benchmark.com/workshops/mai/2021/}
}}

\section{Introduction}

\begin{figure*}[t!]
\centering
\setlength{\tabcolsep}{1pt}
\resizebox{\linewidth}{!}
{
\begin{tabular}{cccc}
\scriptsize{Sony IMX586 RAW -- Visualized}\normalsize & \scriptsize{Sony IMX586 -- MediaTek ISP}\normalsize & \scriptsize{Fujifilm Camera}\normalsize\\
    \includegraphics[width=0.33\linewidth]{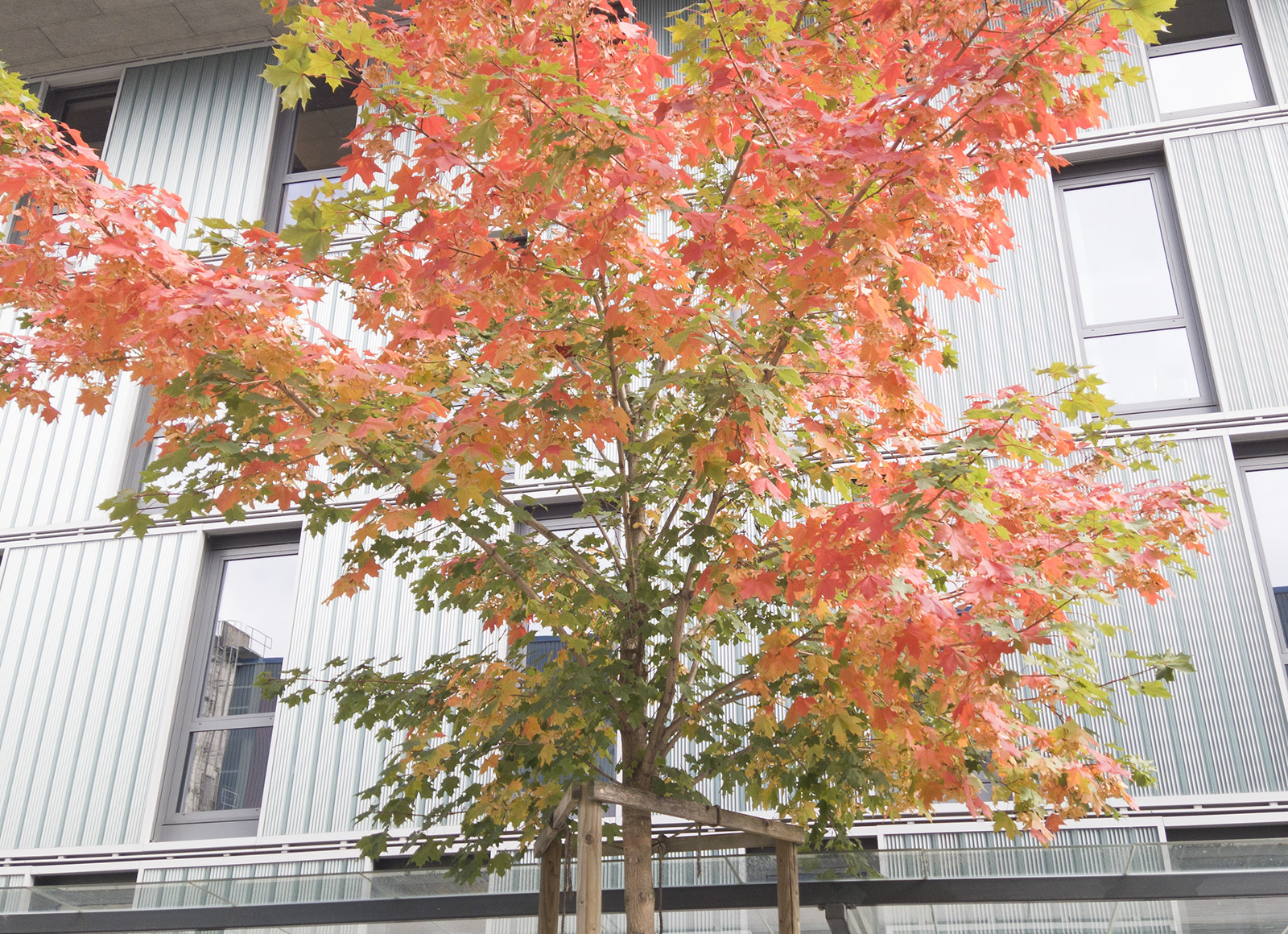}&
    \includegraphics[width=0.33\linewidth]{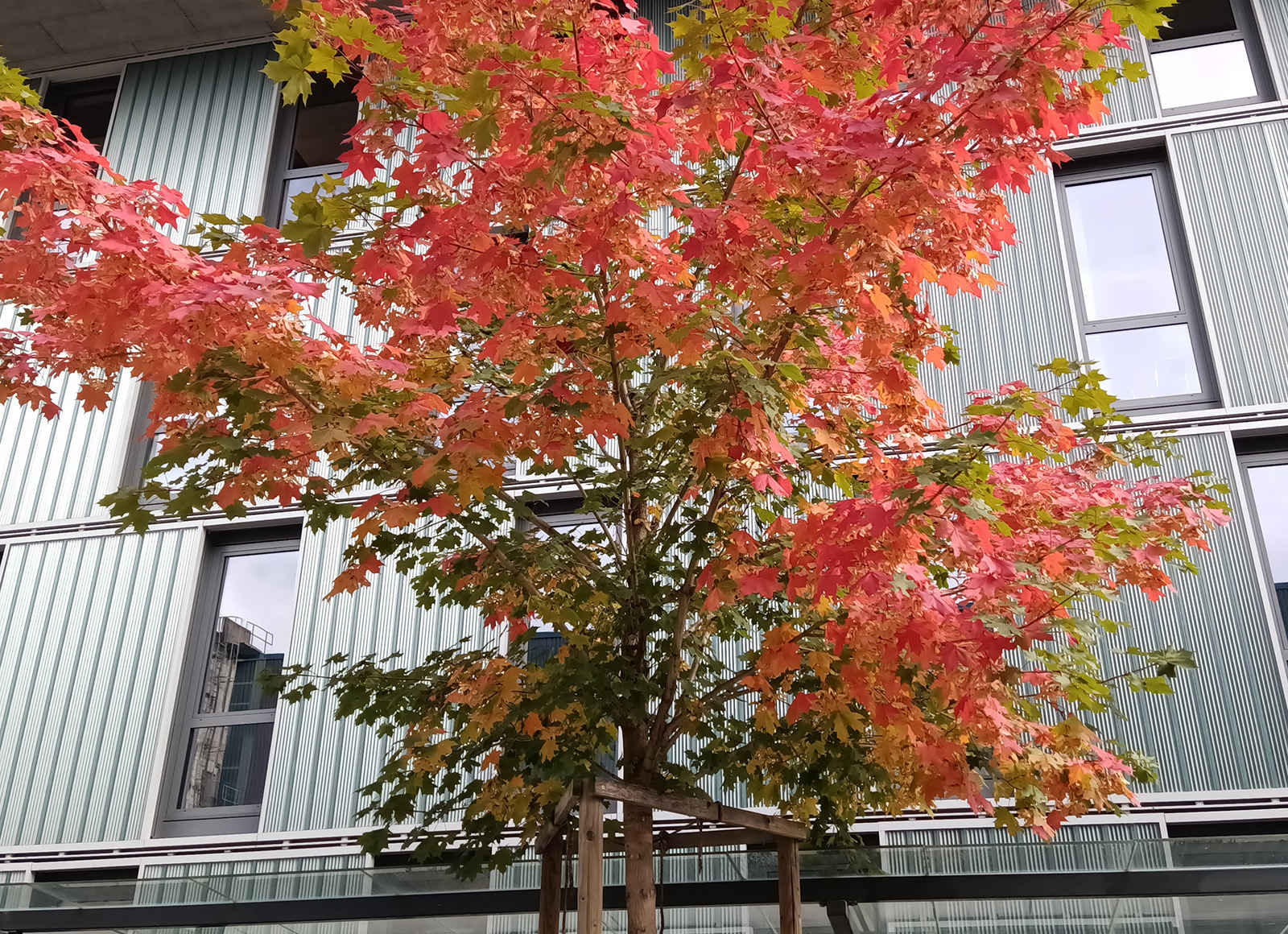}&
    \includegraphics[width=0.33\linewidth]{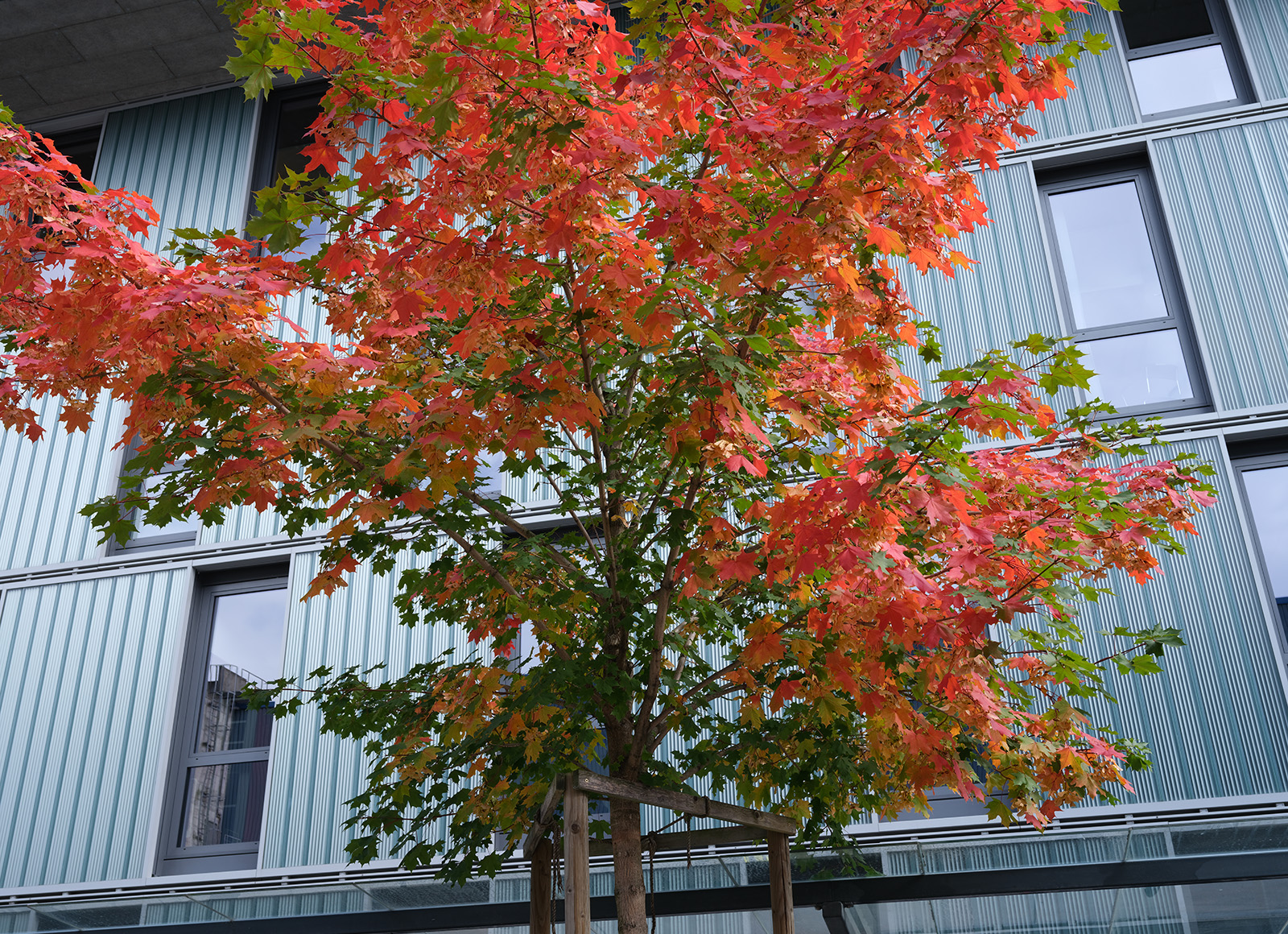}
\end{tabular}
}
\vspace{0.2cm}
\caption{A sample set of images from the collected dataset.  From left to right:  the original RAW image visualized with Photoshop's built-in raw image processing engine, RGB image obtained with MediaTek's built-in ISP system, and the target Fujifilm photo.}
\label{fig:example_photos}
\end{figure*}

While the quality of modern smartphone cameras increases gradually, many major improvements are currently coming from advanced image processing algorithms used, \eg, to perform noise suppression, accurate color reconstruction or high dynamic range processing. Though the image enhancement task can be efficiently solved with deep learning-based approaches, the biggest challenge here comes from getting the appropriate training data and, in particular, the high-quality ground truth images. The problem of end-to-end mobile photo quality enhancement was first addressed in~\cite{ignatov2017dslr,ignatov2018wespe}, where the authors proposed to enhance all aspects of low-quality smartphone photos by mapping them to superior-quality images obtained with a high-end reflex camera. The collected DPED dataset was later used in many subsequent competitions~\cite{ignatov2018pirm,ignatov2019ntire} and works~\cite{vu2018fast,lugmayr2019unsupervised,de2018fast,hui2018perception,huang2018range,liu2018deep} that have significantly improved the results on this problem. While the proposed methods were quite efficient, they worked with the data produced by smartphones' built-in ISPs, thus a significant part of information present in the original sensor data was irrecoverably lost after applying many image processing steps. To address this problem, in~\cite{ignatov2020replacing} the authors proposed to work directly with the original RAW Bayer sensor data and learn all ISP steps with a single deep neural network. The experiments conducted on the collected \textit{Zurich RAW to RGB} dataset containing RAW-RGB image pairs captured by a mobile camera sensor and a high-end DSLR camera demonstrated that the proposed solution was able to get to the level of commercial ISP of the Huawei P20 cameraphone, while these results were later improved in~\cite{ignatov2020aim,dai2020awnet,silva2020deep,kim2020pynet,ignatov2019aim}. In this challenge, we take one step further in solving this problem by using more advanced data and by putting additional efficiency-related constraints on the developed solutions.

When it comes to the deployment of AI-based solutions on mobile devices, one needs to take care of the particularities of mobile NPUs and DSPs to design an efficient model. An extensive overview of smartphone AI acceleration hardware and its performance is provided in~\cite{ignatov2019ai,ignatov2018ai}. According to the results reported in these papers, the latest mobile NPUs are already approaching the results of mid-range desktop GPUs released not long ago. However, there are still two major issues that prevent a straightforward deployment of neural networks on mobile devices: a restricted amount of RAM, and limited and not always efficient support for many common deep learning layers and operators. These two problems make it impossible to process high-resolution data with standard NN models, thus requiring a careful adaptation of each architecture to the restrictions of mobile AI hardware. Such optimizations can include network pruning and compression~\cite{chiang2020deploying,ignatov2020rendering,li2019learning,liu2019metapruning,obukhov2020t}, 16-bit / 8-bit~\cite{chiang2020deploying,jain2019trained,jacob2018quantization,yang2019quantization} and low-bit~\cite{cai2020zeroq,uhlich2019mixed,ignatov2020controlling,liu2018bi} quantization, device- or NPU-specific adaptations, platform-aware neural architecture search~\cite{howard2019searching,tan2019mnasnet,wu2019fbnet,wan2020fbnetv2}, \etc.

While many challenges and works targeted at efficient deep learning models have been proposed recently, the evaluation of the obtained solutions is generally performed on desktop CPUs and GPUs, making the developed solutions not practical due to the above-mentioned issues. To address this problem, we introduce the first \textit{Mobile AI Workshop and Challenges}, where all deep learning solutions are developed for and evaluated on real mobile devices. In this competition, the participating teams were provided with a new ISP dataset consisting of RAW-RGB image pairs captured with the Sony IMX586 mobile sensor and a professional 102-megapixel Fujifilm camera, and were developing an end-to-end deep learning solution for the learned ISP task. Within the challenge, the participants were evaluating the runtime and tuning their models on the MediaTek Dimensity 1000+ platform featuring a dedicated AI Processing Unit (APU) that can accelerate floating-point and quantized neural networks. The final score of each submitted solution was based on the runtime and fidelity results, thus balancing between the image reconstruction quality and efficiency of the proposed model. Finally, all developed solutions are fully compatible with the TensorFlow Lite framework~\cite{TensorFlowLite2021}, thus can be deployed and accelerated on any mobile platform providing AI acceleration through the Android Neural Networks API (NNAPI)~\cite{NNAPI2021} or custom TFLite delegates~\cite{TFLiteDelegates2021}.

\smallskip

This challenge is a part of the \textit{MAI 2021 Workshop and Challenges} consisting of the following competitions:

\small

\begin{itemize}
\item Learned Smartphone ISP on Mobile NPUs
\item Real Image Denoising on Mobile GPUs~\cite{ignatov2021fastDenoising}
\item Quantized Image Super-Resolution on Mobile NPUs~\cite{ignatov2021real}
\item Real-Time Video Super-Resolution on Mobile GPUs~\cite{romero2021real}
\item Single-Image Depth Estimation on Mobile Devices~\cite{ignatov2021fastDepth}
\item Quantized Camera Scene Detection on Smartphones~\cite{ignatov2021fastSceneDetection}
\item High Dynamic Range Image Processing on Mobile NPUs
\end{itemize}

\normalsize

\noindent The results obtained in the other competitions and the description of the proposed solutions can be found in the corresponding challenge report papers.


\begin{figure*}[t!]
\centering
\setlength{\tabcolsep}{1pt}
\resizebox{0.96\linewidth}{!}
{
\includegraphics[width=1.0\linewidth]{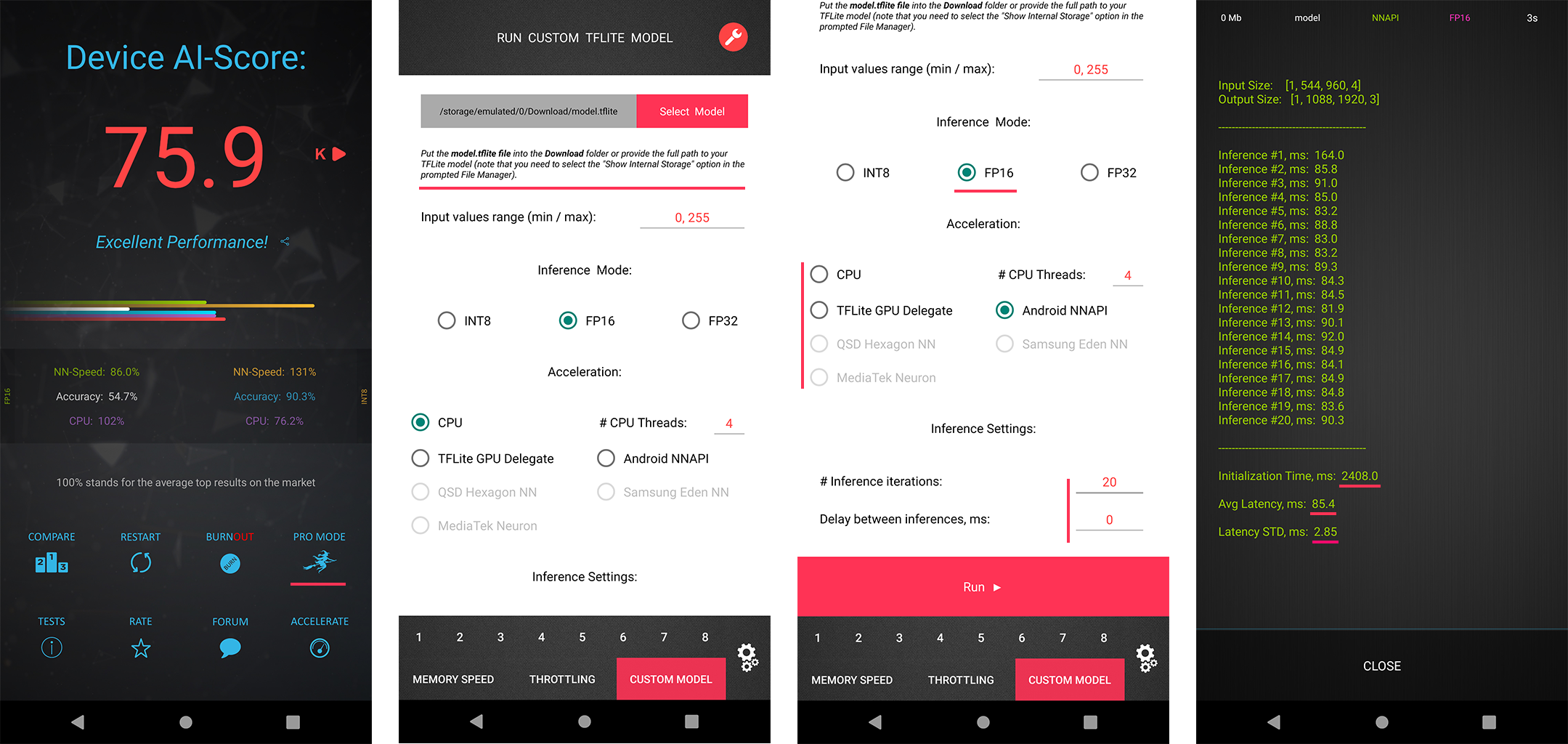}
}
\vspace{0.2cm}
\caption{Loading and running custom TensorFlow Lite models with AI Benchmark application. The currently supported acceleration options include Android NNAPI, TFLite GPU, Hexagon NN, Samsung Eden and MediaTek Neuron delegates as well as CPU inference through TFLite or XNNPACK backends. The latest app version can be downloaded at \url{https://ai-benchmark.com/download}.}
\label{fig:ai_benchmark_custom}
\end{figure*}

\section{Challenge}

To develop an efficient and practical solution for mobile-related tasks, one needs the following major components:

\begin{enumerate}
\item A high-quality and large-scale dataset that can be used to train and evaluate the solution on real (not synthetically generated) data;
\item An efficient way to check the runtime and debug the model locally without any constraints;
\item An ability to regularly test the runtime of the designed neural network on the target mobile platform or device.
\end{enumerate}

This challenge addresses all the above issues. Real training data, tools, and runtime evaluation options provided to the challenge participants are described in the next sections.

\subsection{Dataset}

To handle the problem of image translation from the original RAW photos captured with modern mobile camera sensors to superior quality images achieved by professional full-frame or medium format cameras, a large-scale real-world dataset containing RAW-RGB image pairs was collected. The dataset consists of photos taken in the wild synchronously by a 102-MP Fujifilm medium format camera and the Sony IMX586 Quad Bayer mobile sensor shooting RAW images. The photos were taken during the daytime in a wide variety of places and various illumination and weather conditions. The photos were captured in automatic mode, and the default settings were used for both cameras throughout the whole collection procedure. An example set of collected images can be seen in Fig.~\ref{fig:example_photos}.

Since the captured RAW-RGB image pairs are not perfectly aligned, they were matched using an advanced dense correspondence algorithm~\cite{truong2021learning}, and then smaller patches of size 256$\times$256 px were extracted. The participants were provided with around 24 thousand training RAW-RGB image pairs (of size 256$\times$256$\times$1 and 256$\times$256$\times$3, respectively). It should be mentioned that all alignment operations were performed on RGB Fujifilm images only, therefore RAW photos from the Sony sensor remained unmodified. A comprehensive tutorial demonstrating how to work with the data and how to train a baseline PUNET model on the provided images was additionally released to the participants: \url{https://github.com/MediaTek-NeuroPilot/mai21-learned-smartphone-isp}.

\begin{table*}[t!]
\centering
\resizebox{\linewidth}{!}
{
\begin{tabular}{l|c|cc|cc|c|c}
\hline
Team \, & \, Author \, & \, Framework \, & \, Model Size, KB \, & \, PSNR$\uparrow$ \, & \, SSIM$\uparrow$ & \, Runtime, ms $\downarrow$ \, & \, Final Score \\
\hline
\hline
dh\_isp & \, xushusong001 \, & PyTorch / TensorFlow & 21 & 23.2 & 0.8467 & \textBF{61} & \textBF{25.98} \\
AIISP & AIISP & TensorFlow & 123 & \textBF{23.73} & 0.8487 & 90.8 & \textBF{25.91} \\
\rowcolor{grayhighlight} Tuned U-Net & Baseline & PyTorch / TensorFlow & 13313 & 23.30 & 0.8395 & 78 & 25.74 \\
ENERZAi Research & Minsu.Kwon & TensorFlow & 9 & 22.97 & 0.8392 & 65 & 25.67 \\
isp\_forever & LearnedSmartphoneISP & PyTorch & 175 & 22.78 & 0.8472 & 77 & 25.24 \\
NOAHTCV & noahtcv & TensorFlow & 244 & 23.08 & 0.8237 & 94.5 & 25.19 \\
ACVLab & jesse1029 & TensorFlow & 7 & 22.03 & 0.8217 & 76.3 & 24.5 \\
CVML & vishalchudasama & TensorFlow & 76 & 22.84 & 0.8379 & 167 & 23.5 \\
\textit{ENERZAi Research $^*$} & \textit{jaeyon} & PyTorch / TensorFlow & 11 & 23.41 & \textBF{0.8534} & 231 & 23.39 \\
\rowcolor{grayhighlight} EdS & Etienne & TensorFlow & 1017 & 23.23 & 0.8481 & 1861 & 22.4 \\
\end{tabular}
}
\vspace{2.6mm}
\caption{\small{Mobile AI 2021 smartphone ISP challenge results and final rankings. The runtime values were obtained on Full HD (1920$\times$1088) images. Teams \textit{dh\_isp} and \textit{AIISP} are the challenge winners. \textit{Tuned U-Net} corresponds to a baseline U-Net model~\cite{ronneberger2015u} tuned specifically for the target Dimensity 1000+ platform. Team \textit{EdS} was ranked second in the PIRM 2018 Image Enhancement Challenge~\cite{ignatov2018pirm}, its results on this task are provided for the reference. $^*$~The second solution from \textit{ENERZAi Research} team did not participate in the official test phase, its scores are shown for general information only.}}
\label{tab:results}
\end{table*}

\subsection{Local Runtime Evaluation}

When developing AI solutions for mobile devices, it is vital to be able to test the designed models and debug all emerging issues locally on available devices. For this, the participants were provided with the \textit{AI Benchmark} application~\cite{ignatov2018ai,ignatov2019ai} that allows to load any custom TensorFlow Lite model and run it on any Android device with all supported acceleration options. This tool contains the latest versions of \textit{Android NNAPI, TFLite GPU, Hexagon NN, Samsung Eden} and \textit{MediaTek Neuron} delegates, therefore supporting all current mobile platforms and providing the users with the ability to execute neural networks on smartphone NPUs, APUs, DSPs, GPUs and CPUs.

\smallskip

To load and run a custom TensorFlow Lite model, one needs to follow the next steps:

\begin{enumerate}
\setlength\itemsep{0mm}
\item Download AI Benchmark from the official website\footnote{\url{https://ai-benchmark.com/download}} or from the Google Play\footnote{\url{https://play.google.com/store/apps/details?id=org.benchmark.demo}} and run its standard tests.
\item After the end of the tests, enter the \textit{PRO Mode} and select the \textit{Custom Model} tab there.
\item Rename the exported TFLite model to \textit{model.tflite} and put it into the \textit{Download} folder of the device.
\item Select mode type \textit{(INT8, FP16, or FP32)}, the desired acceleration/inference options and run the model.
\end{enumerate}

\noindent These steps are also illustrated in Fig.~\ref{fig:ai_benchmark_custom}.

\subsection{Runtime Evaluation on the Target Platform}

In this challenge, we use the \textit{MediaTek Dimensity 1000+} SoC as our target runtime evaluation platform. This chipset contains a powerful APU~\cite{lee2018techology} capable of accelerating floating point, INT16 and INT8 models, being ranked first by AI Benchmark at the time of its release~\cite{AIBenchmark202001}. It should be mentioned that FP16/INT16 inference support is essential for this task as raw Bayer data has a high dynamic range (10- to 14-bit images depending on the camera sensor model).

Within the challenge, the participants were able to upload their TFLite models to the runtime validation server connected to a real device and get instantaneous feedback: the runtime of their solution on the Dimensity 1000+ APU or a detailed error log if the model contains some incompatible operations (ops). The models were parsed and accelerated using MediaTek Neuron delegate\footnote{\url{https://github.com/MediaTek-NeuroPilot/tflite-neuron-delegate}}. The same setup was also used for the final runtime evaluation. The participants were additionally provided with a detailed model optimization guideline demonstrating the restrictions and the most efficient setups for each supported TFLite op.

\subsection{Challenge Phases}

The challenge consisted of the following phases:

\vspace{-0.8mm}
\begin{enumerate}
\item[I.] \textit{Development:} the participants get access to the data and AI Benchmark app, and are able to train the models and evaluate their runtime locally;
\item[II.] \textit{Validation:} the participants can upload their models to the remote server to check the fidelity scores on the validation dataset, to get the runtime on the target platform, and to compare their results on the validation leaderboard;
\item[III.] \textit{Testing:} the participants submit their final results, codes, TensorFlow Lite models, and factsheets.
\end{enumerate}
\vspace{-0.8mm}

\subsection{Scoring System}

All solutions were evaluated using the following metrics:

\vspace{-0.8mm}
\begin{itemize}
\setlength\itemsep{-0.2mm}
\item Peak Signal-to-Noise Ratio (PSNR) measuring fidelity score,
\item Structural Similarity Index Measure (SSIM), a proxy for perceptual score,
\item The runtime on the target Dimensity 1000+ platform.
\end{itemize}
\vspace{-0.8mm}

The score of each final submission was evaluated based on the next formula:
\begin{equation*}
\text{Final Score} = \text{PSNR} + \alpha \cdot (0.2 - clip(\text{runtime})),
\end{equation*}
where:

\begin{equation*}
\alpha = \begin{cases}
      20, & \text{if}\, \text{runtime} \leq 0.2 \\
      0.5, & \text{otherwise}
    \end{cases},
\end{equation*}

\begin{equation*}
clip = min(max(\text{runtime}, 0.03), 5).
\end{equation*}

During the final challenge phase, the participants did not have access to the test dataset. Instead, they had to submit their final TensorFlow Lite models that were subsequently used by the challenge organizers to check both the runtime and the fidelity results of each submission under identical conditions. This approach solved all the issues related to model overfitting, reproducibility of the results, and consistency of the obtained runtime/accuracy values.

\section{Challenge Results}

From the above 190 registered participants, 9 teams entered the final phase and submitted valid results, TFLite models, codes, executables, and factsheets. Table~\ref{tab:results} summarizes the final challenge results and  reports PSNR, SSIM, and runtime numbers for each submitted solution on the final test dataset on the target evaluation platform. The proposed methods are described in Section~\ref{sec:solutions}, and the team members and affiliations are listed in Appendix~\ref{sec:apd:team}.


\subsection{Results and Discussion}

All submitted solutions demonstrated a very high efficiency: the majority of models are able to process one Full HD (1920$\times$1088 px) image under 100 ms on the target MediaTek APU. Teams \textit{dh\_isp} and \textit{AIISP} are the winners of this challenge, achieving the best runtime and fidelity results on this task, respectively. These solutions are following two absolutely different approaches. \textit{dh\_isp} is using an extremely shallow 3-layer \textit{FSRCNN}~\cite{dong2016accelerating}-inspired model with one pixel-shuffle block, and is processing the input image at the original scale. The size of this model is only 21 KB, and it is able to achieve an impressive 16 FPS on the Dimensity 1000+ SoC. In contrast, the solution proposed by \textit{AIISP} is downsampling the input data and applying a number of convolutional layers and several sophisticated attention blocks to get high fidelity results, while also demonstrating more than 10 FPS on the target platform.

\textit{Tuned U-Net} is a solid U-Net baseline with several hardware-driven adaptations designed to demonstrate the performance that can be achieved with a common APU-aware tuned deep learning architecture. It is also showing that the weight and the number of layers do not necessarily play key roles in model efficiency if the majority of processing is happening at lower scales. While its size is more than 1000 times larger compared to the model proposed by \textit{ENERZAi Research}, it demonstrates comparable runtime results on the same hardware. The tremendous model size reduction in the latter case was achieved by using an efficient knowledge transfer approach consisting of the joint training of two (tiny and large) models sharing the same feature extraction block. Another interesting approach was proposed by \textit{NOAHTCV} which model is processing chroma and texture information separately.

In the final ranking table, one can also find the results of team \textit{EdS} that was ranked second in the \textit{PIRM 2018 Image Enhancement Challenge}~\cite{ignatov2018pirm}. This model was deliberately not optimized for the target platform to demonstrate the importance of such fine-tuning. While its fidelity scores are still high, showing the second-best PSNR result, it requires almost 2 seconds to process one image on the Dimensity 1000+ platform. The reason for this is quite straightforward: it is using several ops not yet adequately supported by NNAPI (despite claimed as officially supported). By removing or replacing these ops, the runtime of this model improves by more than 10 times, while the corresponding difference on desktop CPUs / GPUs is less than 10\%. This example explicitly shows that the runtime values obtained on common deep learning hardware are not representative when it comes to model deployment on mobile AI silicon: even solutions that might seem to be very efficient can struggle significantly due to the specific constraints of smartphone AI acceleration hardware and frameworks. This makes deep learning development for mobile devices so challenging, though the results obtained in this competition demonstrate that one can get a very efficient model when taking the above aspects into account.

\section{Challenge Methods}
\label{sec:solutions}

\noindent This section describes solutions submitted by all teams participating in the final stage of the MAI 2021 Learned Smartphone ISP challenge.

\subsection{dh\_isp}

\begin{figure}[h!]
\centering
\resizebox{1.0\linewidth}{!}
{
\includegraphics[width=1.0\linewidth]{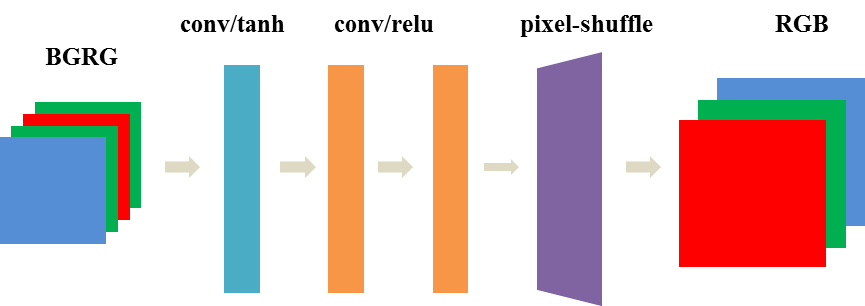}
}
\caption{\small{Smallnet architecture proposed by team dh\_isp.}}
\label{fig:dhisp}
\end{figure}

Team dh\_isp proposed a very compact Smallnet architecture illustrated in Fig.~\ref{fig:dhisp} that consists of three convolutional and one pixel-shuffle layer. Each convolutional layer is using 3$\times$3 kernels and has 16, 16, and 12 channels, respectively. \textit{Tanh} activation function is used after the first layer, while the other ones are followed by the \textit{ReLU} function. The authors especially emphasize the role of the \textit{Tanh} and pixel-shuffle ops in getting high fidelity results on this task.

The model was first trained with $L_1$ loss only, and then fine-tuned with a combination of $L_1$ and perceptual-based \mbox{\textit{VGG-19}} loss functions. The network parameters were optimized with the Adam algorithm using a batch size of 4 and a learning rate of $1e-4$ that was decreased within the training.

\subsection{AIISP}

\begin{figure*}[t!]
\centering
\resizebox{0.8\linewidth}{!}
{
\includegraphics[width=1.0\linewidth]{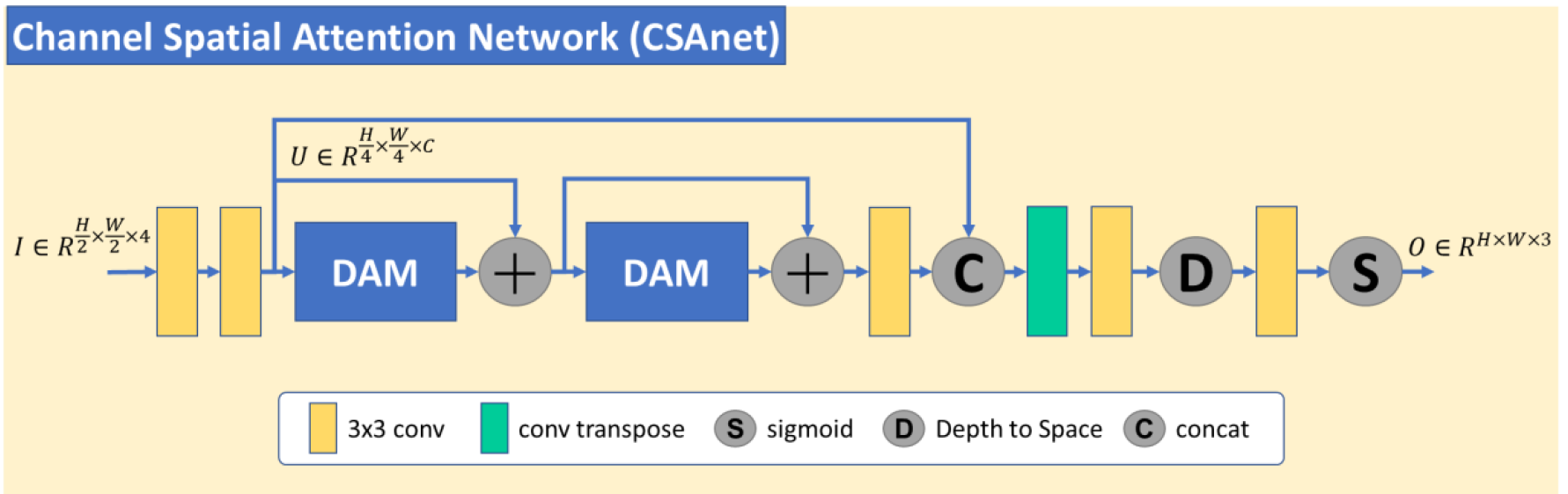}
}
\caption{\small{CSANet model from team AIISP with several attention blocks.}}
\label{fig:AIISP1}
\end{figure*}

\begin{figure}[h!]
\centering
\resizebox{0.8\linewidth}{!}
{
\includegraphics[width=1.0\linewidth]{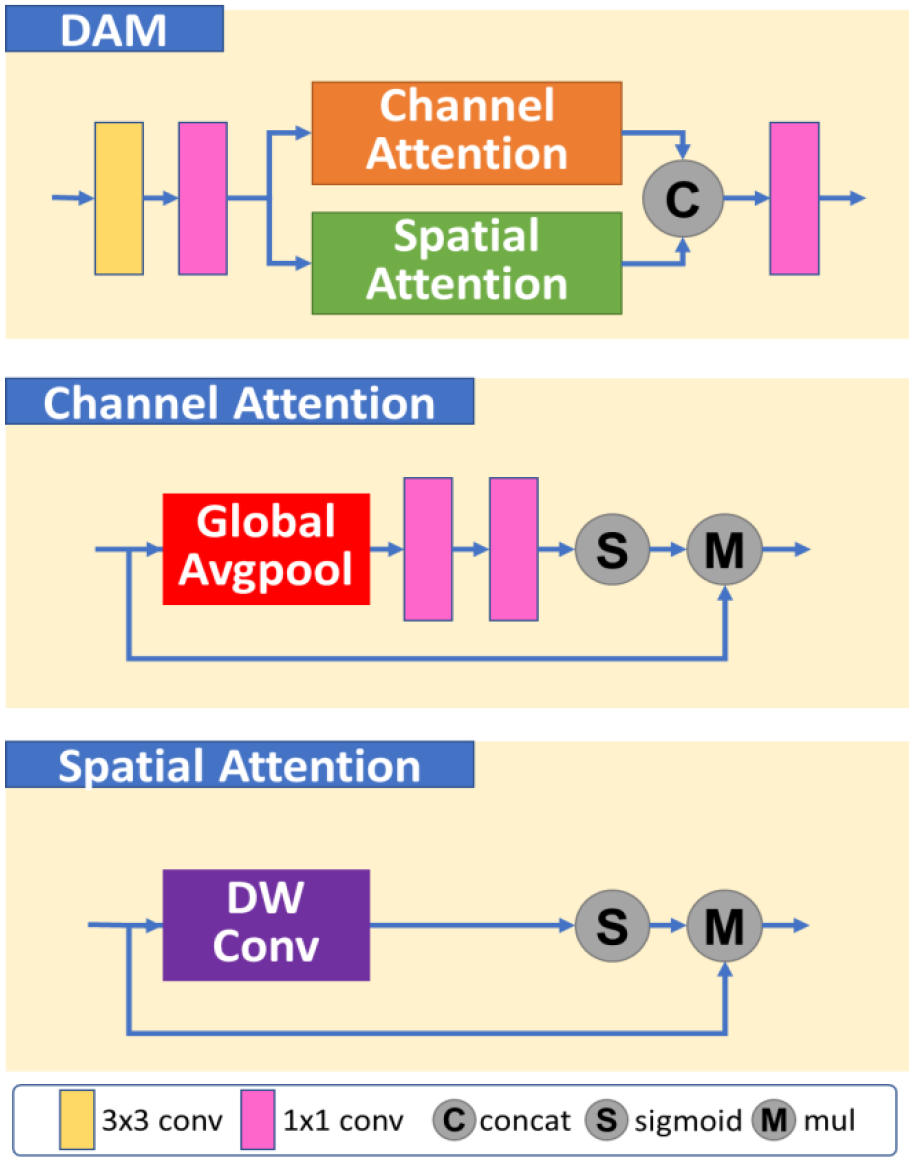}
}
\caption{\small{The structure of double attention module (DAM), spatial attention and channel attention blocks.}}
\label{fig:AIISP2}
\end{figure}

The authors proposed a Channel Spacial Attention Network (Fig.~\ref{fig:AIISP1}) that achieved the best fidelity results in this challenge. The architecture of this model consists of the following three main parts. In the first part, two convolutional blocks with \textit{ReLU} activations are used to perform feature extraction and downsize the input RAW data. After that, a series of processing blocks are cascaded. The middle double attention modules (DAM) with skip connections are mainly designed to enhance the spatial dependencies and to highlight the prominent objects in the feature maps. These skip connections are used not only to avoid the vanishing gradient problem but also to keep the similarities between the learned feature maps from different blocks. The last part of the network uses transposed convolution and depth-to-space modules to upscale the feature maps to their target size. Finally, a conventional convolution followed by the sigmoid activation function restores the output RGB image.

The sub-network structure of DAM is shown in Fig.~\ref{fig:AIISP2}. Given the feature maps obtained after applying two convolutions, DAM performs feature recalibration by using two attention mechanisms: spatial attention (SA) and channel attention (CA). The results of these concatenated attentions are then followed by a 1$\times$1 convolutional layer to yield an adaptive feature refinement. The spatial attention module is designed to learn spatial dependencies in the feature maps. In order to have a distant vision over these maps, a depth-wise dilated convolution with a 5$\times$5 kernel is used to extract the information. The output of this module is multiplied with the corresponding input feature map to get the final result. Channel attention block uses squeeze-and-excite operations to learn the inter-channel relationship between the feature maps. The squeeze operation is implemented by computing the mean values over individual feature maps. The excite operation is composed of two 1$\times$1 convolution layers with different channel sizes and activations (\textit{ReLU} and sigmoid, respectively) and re-calibrates the squeeze output. The output of the module is also obtained by elemental-wise multiplication of the input feature maps and the calibrated descriptor. A more detailed description of the CSANet architecture is provided in~\cite{hsyu2021CSANet}.

The model is trained with a combination of the \textit{Charbonnier} loss function (used to approximate $L_1$ loss), perceptual \textit{VGG-19} and \textit{SSIM} losses. The weights of the model are optimized using Adam for 100K iterations with a learning rate of $5e-4$ decreased to $1e-5$ throughout the training. A batch size of 100 was used, the training data was augmented by random horizontal flipping.

\subsection{Tuned U-Net Baseline}

A U-Net~\cite{ronneberger2015u} based model was developed to get an effective baseline for this challenge. This model follows the standard U-Net architecture with skip connections, and introduces several hardware-specific adaptations for the target platform such as a reduced number of feature maps, modified convolutional filter sizes, and activation functions, and additional skip connections used to maintain a reasonable accuracy. The model was trained with a combination of \textit{MSE} and \textit{SSIM} loss functions using Adam optimizer with a learning rate of $1e-4$.

\subsection{ENERZAi Research}

\begin{figure}[h!]
\centering
\resizebox{1.0\linewidth}{!}
{
\includegraphics[width=1.0\linewidth]{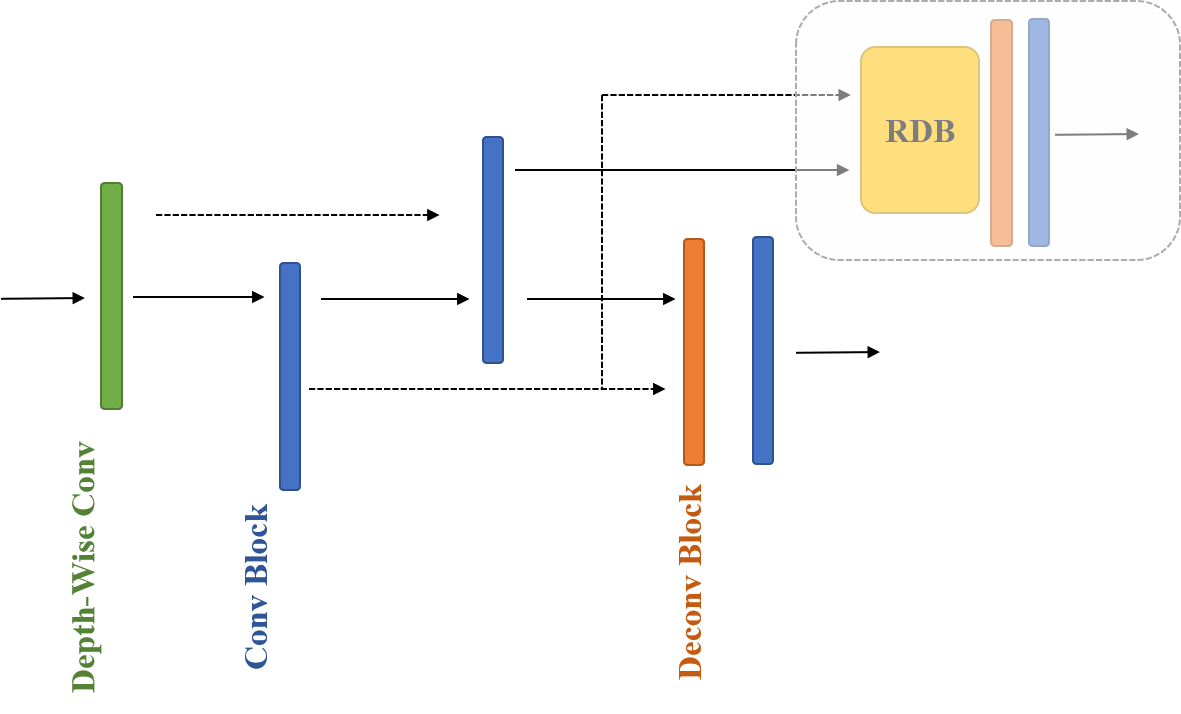}
}
\caption{\small{The model architecture proposed by ENERZAi Research team. Semitransparent residual dense block belongs to the super-network and is detached after training.}}
\label{fig:ENERZAi}
\end{figure}

\begin{figure}[h!]
\centering
\resizebox{1.0\linewidth}{!}
{
\includegraphics[width=1.0\linewidth]{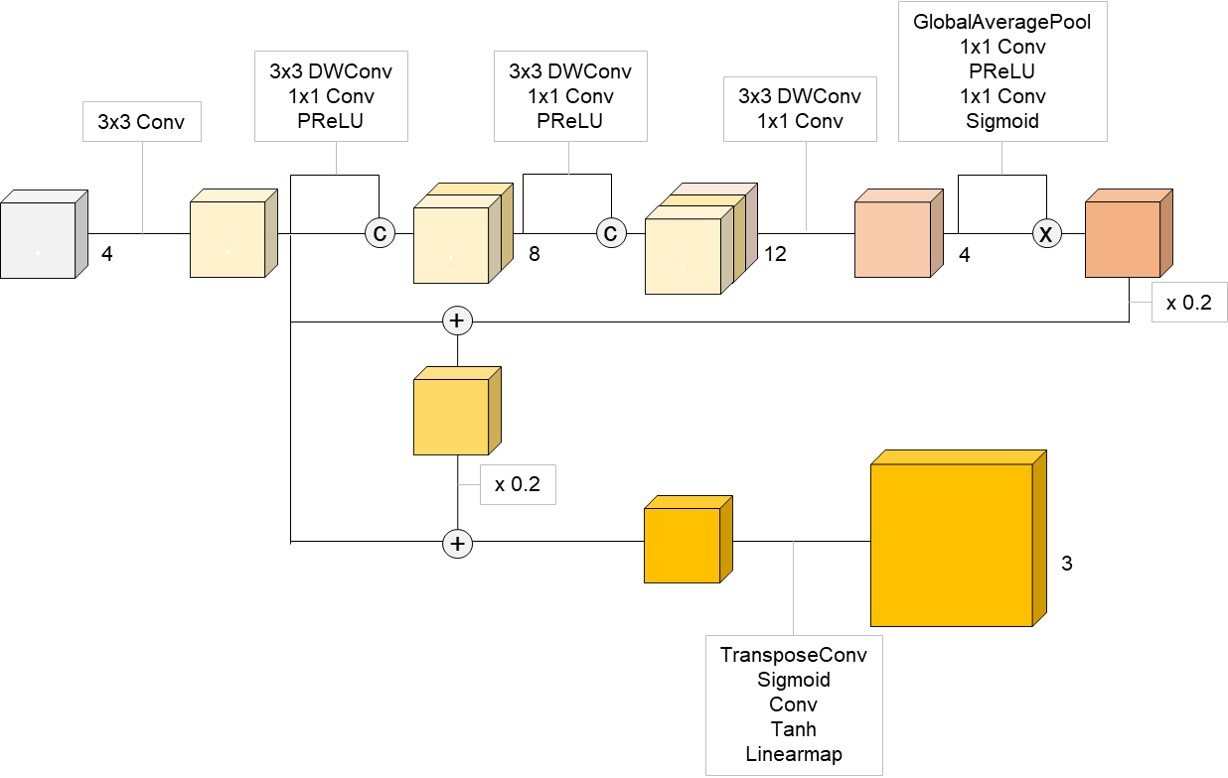}
}
\caption{\small{ESRGAN-based architecture with additional attention block proposed by ENERZAi Research.}}
\label{fig:ENERZAi2}
\end{figure}

The solution proposed by ENERZAi Research is inspired by the \textit{Once-for-All} approach~\cite{cai2019once} and consists of two models: one super-network and one sub-network. They both share the same Dense-Net-like module, and the difference comes from their top layers: the sub-network has one deconvolution, convolution, and sigmoid layers, while the super-network additionally contains several residual dense blocks as shown in Fig.~\ref{fig:ENERZAi}. Both models are first trained jointly using a combination of the \textit{Charbonnier} and \textit{MS-SSIM} loss functions. The super-network is then detached after the PSNR score goes above a predefined  threshold, and the sub-net is further fine-tuned alone. The model was trained using Adam optimizer with a batch size of 4 and a learning rate of $1e-3$.


The second model proposed by this team (which did not officially participate in the final test phase) is demonstrated in Fig.~\ref{fig:ENERZAi2}. It follows the ESRGAN architecture~\cite{wang2018esrgan} and has a shallow feature extractor and several DenseNet-based residual blocks with separable convolutions followed by a transpose convolution layer. The authors also used an additional channel attention block to boost the fidelity results at the expense of a very slight speed degradation. To choose the most appropriate activation function, the authors applied NAS technique that resulted in selecting the \textit{PReLU} activations. The model was trained with a combination of the \textit{MS-SSIM} and $L_1$ losses. It should be also mentioned that the original model was implemented and trained using PyTorch. To avoid the problems related to inefficient PyTorch-to-TensorFlow conversion, the authors developed their own scripts translating the original model architecture and weights to TensorFlow, and then converted the obtained network to TFLite.

\subsection{isp\_forever}

\begin{figure}[h!]
\centering
\resizebox{1.0\linewidth}{!}
{
\includegraphics[width=1.0\linewidth]{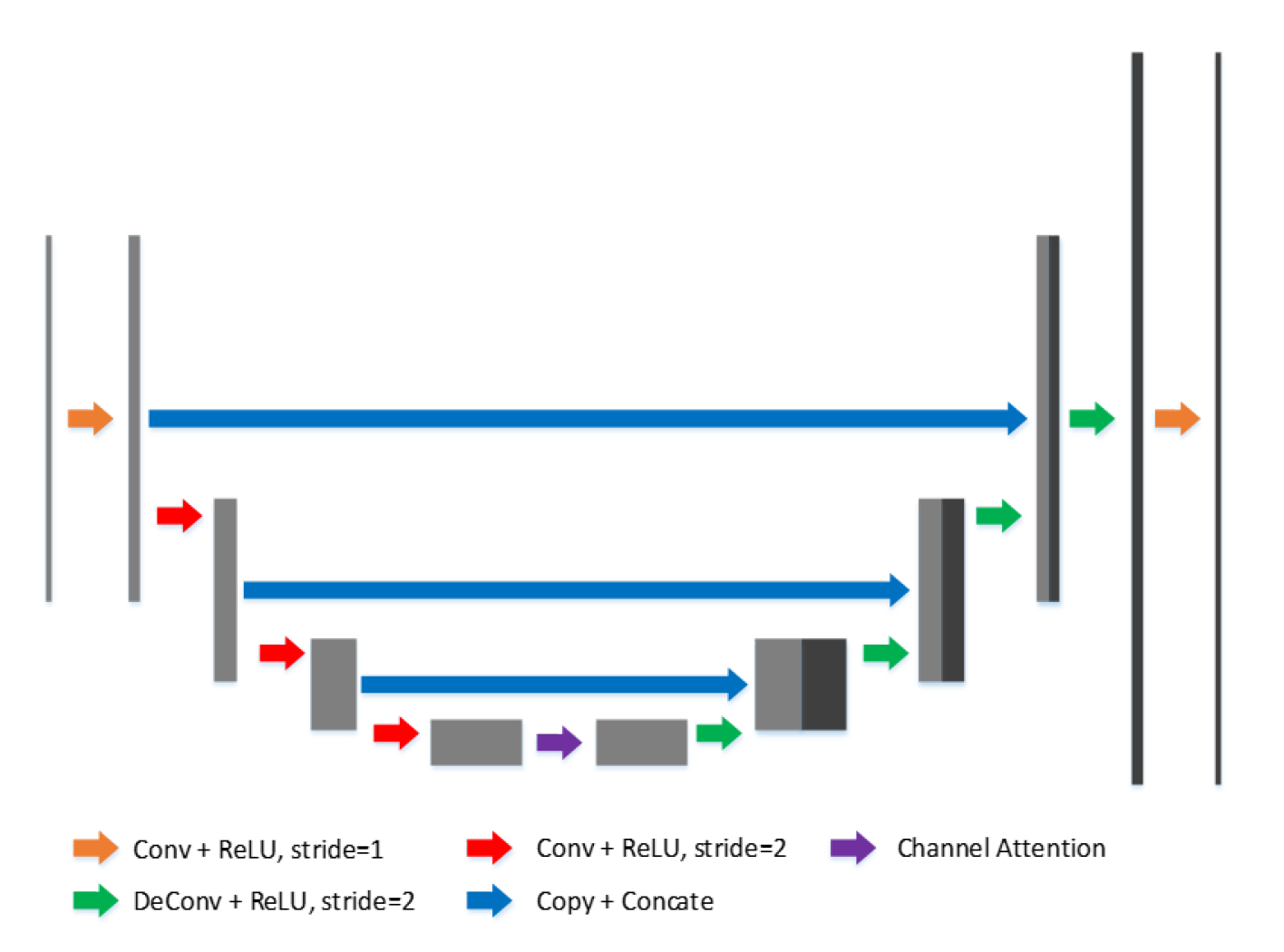}
}
\caption{\small{U-Net based network with a channel attention module from isp\_forever.}}
\label{fig:ispforever}
\end{figure}

Team isp\_forever proposed another truncated U-Net based model for this task that is demonstrated in Fig.~\ref{fig:ispforever}. The authors augmented their network with a channel attention module and trained the entire model with a combination of $L_1$, \textit{SSIM}, and \textit{VGG}-based losses using Adam optimizer with a learning rate of $1e-4$ and a batch size of 16.

\subsection{NOAHTCV}

\begin{figure}[h!]
\centering
\resizebox{1.0\linewidth}{!}
{
\includegraphics[width=1.0\linewidth]{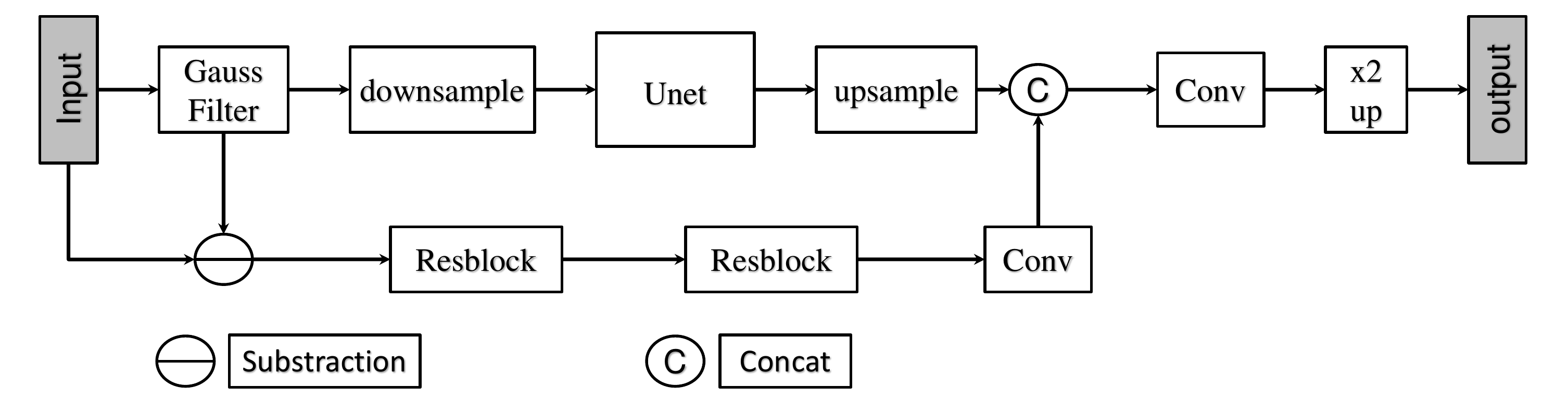}
}
\caption{\small{Network architecture proposed by team NOAHTCV and processing chroma and texture parts separately.}}
\label{fig:NOAHTCV}
\end{figure}

This team proposed to decompose the input image into two parts (Fig.~\ref{fig:NOAHTCV}): chroma part that contains color information, and texture part that includes high-frequency details. The network processes these two parts in separate paths: the first one has a U-Net like architecture and performs patch-level information extraction and tone-mapping, while the second one applies residual blocks for texture enhancement. The outputs from both paths are fused at the end and then upsampled to get the final result. $L_1$ and \textit{SSIM} losses were used to train the network, its parameters were initialized with Xavier and optimized using Adam with a learning rate of $1e-4$.

\subsection{ACVLab}

ACVLab proposed a very compact CNN model with a local fusion block. This block consists of two parts: multiple stacked adaptive weight residual units, termed as RRDB (residual in residual dense block), and a local residual fusion unit (LRFU). The RRDB module can improve the information flow and gradients, while the LRFU module can effectively fuse multi-level residual information in the local fusion block. The model was trained using \textit{VGG}-based, \textit{SSIM}, and \textit{Smooth} $L_1$ losses.

\subsection{CVML}

\begin{figure}[h!]
\centering
\resizebox{1.0\linewidth}{!}
{
\includegraphics[width=1.0\linewidth]{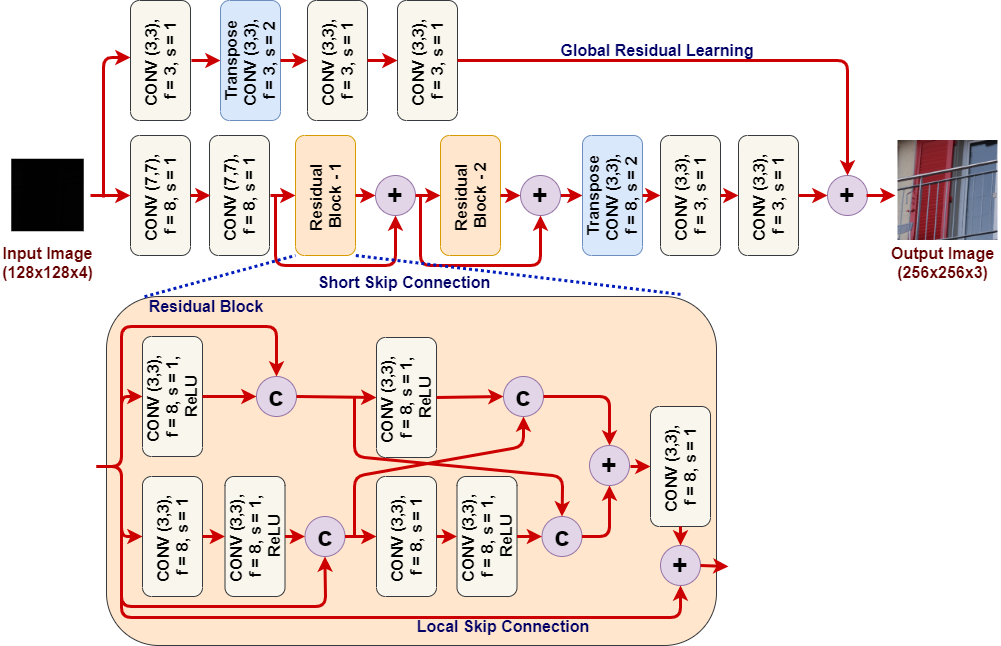}
}
\caption{\small{CNN architecture proposed by CVML team.}}
\label{fig:CVML}
\end{figure}

Figure~\ref{fig:CVML} demonstrates the model proposed by team CVML. This architecture is using residual blocks to extracts a rich set of features from the input data. Transposed convolution layer is used to upsample the final feature maps to the target resolution. To stabilize the training process, a global residual learning strategy was employed that also helped to reduce the color shift effect. The model was trained to minimize the combination of $L_1$ and \textit{SSIM} losses and was optimized using Adam with a learning rate of $1e-4$ for 1M iterations.

\subsection{EdS}

\begin{figure}[h!]
\centering
\resizebox{1.0\linewidth}{!}
{
\includegraphics[width=1.0\linewidth]{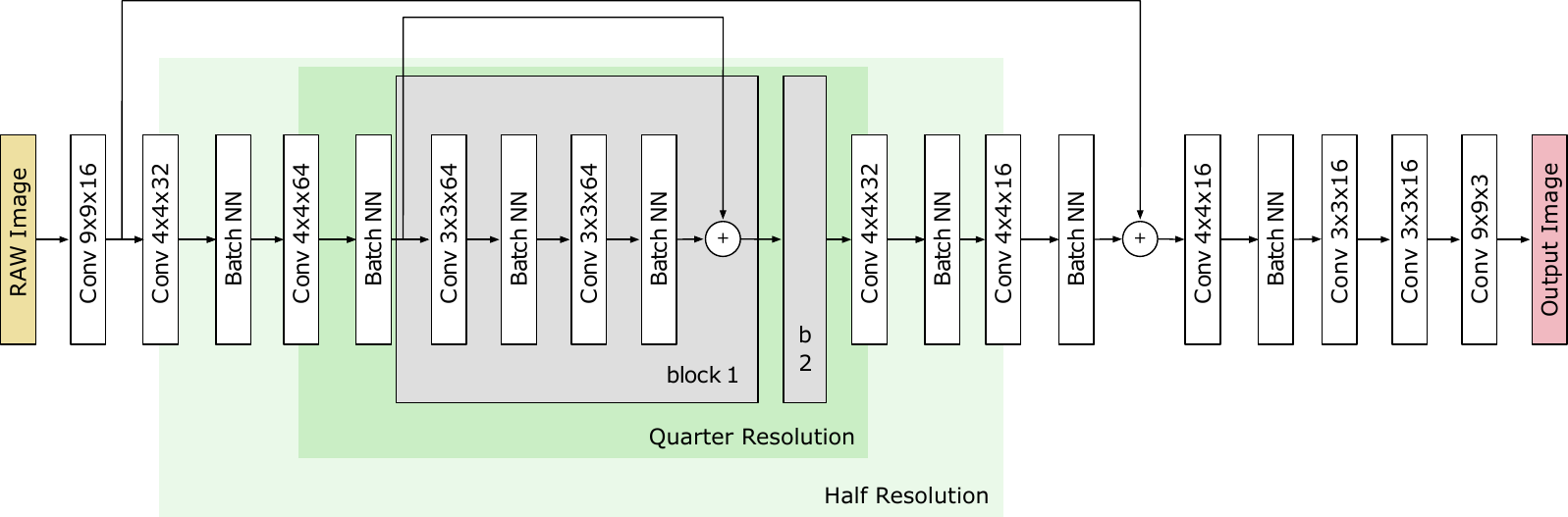}
}
\caption{\small{Residual network proposed by team EdS.}}
\label{fig:EdS}
\end{figure}

EdS proposed a ResNet-based architecture shown in Fig.~\ref{fig:EdS} that was derived from~\cite{ignatov2017dslr}. The main difference consists in using two 4$\times$4 convolutional layers with stride 2 for going into lower-dimensional space, and additional skip connections for faster training. The network was trained for 33K iterations using the same losses and setup as in~\cite{de2018fast}.

\section{Additional Literature}

An overview of the past challenges on mobile-related tasks together with the proposed solutions can be found in the following papers:

\begin{itemize}
\item Learned End-to-End ISP:\, \cite{ignatov2019aim,ignatov2020aim}
\item Perceptual Image Enhancement:\, \cite{ignatov2018pirm,ignatov2019ntire}
\item Bokeh Effect Rendering:\, \cite{ignatov2019aimBokeh,ignatov2020aimBokeh}
\item Image Super-Resolution:\, \cite{ignatov2018pirm,lugmayr2020ntire,cai2019ntire,timofte2018ntire}
\end{itemize}

\section*{Acknowledgements}

We thank MediaTek Inc., AI Witchlabs and ETH Zurich (Computer Vision Lab), the organizers and sponsors of this Mobile AI 2021 challenge.

\appendix
\section{Teams and Affiliations}
\label{sec:apd:team}

\bigskip

\subsection*{Mobile AI 2021 Team}
\noindent\textit{\textbf{Title: }}\\ Mobile AI 2021 Learned Smartphone ISP Challenge\\
\noindent\textit{\textbf{Members:}}\\ Andrey Ignatov$^{1,3}$ \textit{(andrey@vision.ee.ethz.ch)}, Cheng-Ming Chiang$^2$ \textit{(jimmy.chiang@mediatek.com)}, Hsien-Kai Kuo$^2$ \textit{(hsienkai.kuo@mediatek.com)}, Anastasia Sycheva$^1$, Radu Timofte$^{1,3}$  \textit{(radu.timofte@vision.ee.ethz.ch)}, Min-Hung Chen$^2$ \textit{(mh.chen@mediatek.com)}, Man-Yu Lee$^2$ \textit{(my.lee@mediatek.com)}, Yu-Syuan Xu$^2$ \textit{(yu-syuan.xu @mediatek.com)}, Yu Tseng$^2$ \textit{(ulia.tseng@mediatek.com)}\\
\noindent\textit{\textbf{Affiliations: }}\\
$^1$ Computer Vision Lab, ETH Zurich, Switzerland\\
$^2$ MediaTek Inc., Taiwan\\
$^3$ AI Witchlabs, Switzerland\\

\subsection*{dh\_isp}
\noindent\textit{\textbf{Title:}}\\ Smallnet for An End-to-end ISP Pipeline\\
\noindent\textit{\textbf{Members: }}\\ \textit{Shusong Xu (13821177832@163.com)}, Jin Guo\\
\noindent\textit{\textbf{Affiliations: }}\\
Dahua Technology, China\\

\subsection*{AIISP}
\noindent\textit{\textbf{Title:}}\\CSAnet: High Speed Channel Spacial Attention Network for Mobile ISP~\cite{hsyu2021CSANet}\\
\noindent\textit{\textbf{Members:}}\\ \textit{Chao-Hung Chen (ZHChen@itri.org.tw)}, Ming-Chun Hsyu, Wen-Chia Tsai, Chao-Wei Chen\\
\noindent\textit{\textbf{Affiliations: }}\\
Industrial Technology Research Institute (ITRI), Taiwan\\

\subsection*{U-Net Baseline}
\noindent\textit{\textbf{Title:}}\\Platform-aware Tuned U-Net Model\\
\noindent\textit{\textbf{Members:}}\\ \textit{Grigory Malivenko (grigory.malivenko@gmail.com)}\\
\noindent\textit{\textbf{Affiliations: }}\\
Out of competition participation\\

\subsection*{ENERZAi Research}
\noindent\textit{\textbf{Title:}}\\Learning DenseNet by Shrinking Large Network for ISP\\
\noindent\textit{\textbf{Members:}}\\ \textit{Minsu Kwon (minsu.kwon@enerzai.com)}, Myungje Lee, Jaeyoon Yoo, Changbeom Kang, Shinjo Wang\\
\noindent\textit{\textbf{Affiliations: }}\\
ENERZAi, Seoul, Korea \\ \textit{enerzai.com}\\

\subsection*{isp\_forever}
\noindent\textit{\textbf{Title:}}\\A Lightweight U-Net Model for Learned Smartphone ISP\\
\noindent\textit{\textbf{Members:}}\\ \textit{Zheng Shaolong (master5158@163.com)}, Hao Dejun, Xie Fen, Feng Zhuang\\
\noindent\textit{\textbf{Affiliations: }}\\
Dahua Technology, China\\

\subsection*{NOAHTCV}
\noindent\textit{\textbf{Title:}}\\Decomposition Network for AutoISP\\
\noindent\textit{\textbf{Members:}}\\ \textit{Yipeng Ma (mayipeng@huawei.com)}, Jingyang Peng, Tao Wang, Fenglong Song\\
\noindent\textit{\textbf{Affiliations: }}\\
Huawei Noah's Ark Lab, China\\

\subsection*{ACVLab}
\noindent\textit{\textbf{Title:}}\\Lightweight Residual Learning Network for Learned ISP\\
\noindent\textit{\textbf{Members:}}\\ \textit{Chih-Chung Hsu (cchsu@gs.ncku.edu.tw)}, Kwan-Lin Chen, Kwan-Lin Chen, Mei-Hsuang Wu\\
\noindent\textit{\textbf{Affiliations: }}\\
Institute of Data Science, National Cheng Kung University, Taiwan\\

\subsection*{CVML}
\noindent\textit{\textbf{Title:}}\\Mobile compatible Convolution Neural Network for Learned Smartphone ISP\\
\noindent\textit{\textbf{Members:}}\\ \textit{Vishal Chudasama (vishalmchudasama88@gmail.com)}, Kalpesh Prajapati, Heena Patel, Anjali Sarvaiya, Kishor Upla, Kiran Raja, Raghavendra Ramachandra, Christoph Busch\\
\noindent\textit{\textbf{Affiliations: }}\\
Sardar Vallabhbhai National Institute of Technology, India\\

\subsection*{EdS}
\noindent\textit{\textbf{Title:}}\\Fast learned ISP using Resnet Variant with Transposed Convolutions\\
\noindent\textit{\textbf{Members:}}\\ \textit{Etienne de Stoutz (edestoutz@gmail.com)}\\
\noindent\textit{\textbf{Affiliations: }}\\
ETH Zurich, Switzerland\\

{\small
\bibliographystyle{ieee_fullname}

\begin{thebibliography}{10}\itemsep=-1pt

\bibitem{NNAPI2021}
Android Neural~Networks API.
\newblock https://developer.android.com/ndk/guides/neuralnetworks.

\bibitem{AIBenchmark202001}
AI~Benchmark Archive.
\newblock https://bit.ly/32wykta.

\bibitem{cai2019once}
Han Cai, Chuang Gan, Tianzhe Wang, Zhekai Zhang, and Song Han.
\newblock Once-for-all: Train one network and specialize it for efficient
  deployment.
\newblock {\em arXiv preprint arXiv:1908.09791}, 2019.

\bibitem{cai2019ntire}
Jianrui Cai, Shuhang Gu, Radu Timofte, and Lei Zhang.
\newblock Ntire 2019 challenge on real image super-resolution: Methods and
  results.
\newblock In {\em Proceedings of the IEEE/CVF Conference on Computer Vision and
  Pattern Recognition Workshops}, pages 0--0, 2019.

\bibitem{cai2020zeroq}
Yaohui Cai, Zhewei Yao, Zhen Dong, Amir Gholami, Michael~W Mahoney, and Kurt
  Keutzer.
\newblock Zeroq: A novel zero shot quantization framework.
\newblock In {\em Proceedings of the IEEE/CVF Conference on Computer Vision and
  Pattern Recognition}, pages 13169--13178, 2020.

\bibitem{chiang2020deploying}
Cheng-Ming Chiang, Yu Tseng, Yu-Syuan Xu, Hsien-Kai Kuo, Yi-Min Tsai, Guan-Yu
  Chen, Koan-Sin Tan, Wei-Ting Wang, Yu-Chieh Lin, Shou-Yao~Roy Tseng, et~al.
\newblock Deploying image deblurring across mobile devices: A perspective of
  quality and latency.
\newblock In {\em Proceedings of the IEEE/CVF Conference on Computer Vision and
  Pattern Recognition Workshops}, pages 502--503, 2020.

\bibitem{dai2020awnet}
Linhui Dai, Xiaohong Liu, Chengqi Li, and Jun Chen.
\newblock Awnet: Attentive wavelet network for image isp.
\newblock {\em arXiv preprint arXiv:2008.09228}, 2020.

\bibitem{de2018fast}
Etienne de Stoutz, Andrey Ignatov, Nikolay Kobyshev, Radu Timofte, and Luc
  Van~Gool.
\newblock Fast perceptual image enhancement.
\newblock In {\em Proceedings of the European Conference on Computer Vision
  (ECCV) Workshops}, pages 0--0, 2018.

\bibitem{TFLiteDelegates2021}
TensorFlow~Lite delegates.
\newblock https://www.tensorflow.org/lite/performance/delegates.

\bibitem{dong2016accelerating}
Chao Dong, Chen~Change Loy, and Xiaoou Tang.
\newblock Accelerating the super-resolution convolutional neural network.
\newblock In {\em European conference on computer vision}, pages 391--407.
  Springer, 2016.

\bibitem{howard2019searching}
Andrew Howard, Mark Sandler, Grace Chu, Liang-Chieh Chen, Bo Chen, Mingxing
  Tan, Weijun Wang, Yukun Zhu, Ruoming Pang, Vijay Vasudevan, et~al.
\newblock Searching for mobilenetv3.
\newblock In {\em Proceedings of the IEEE/CVF International Conference on
  Computer Vision}, pages 1314--1324, 2019.

\bibitem{hsyu2021CSANet}
Ming-Chun Hsyu, Chih-Wei Liu, Chao-Hung Chen, Chao-Wei Chen, and Wen-Chia Tsai.
\newblock Csanet: High speed channel spatial attention network for mobile isp.
\newblock In {\em Proceedings of the IEEE/CVF Conference on Computer Vision and
  Pattern Recognition Workshops}, pages 0--0, 2021.

\bibitem{huang2018range}
Jie Huang, Pengfei Zhu, Mingrui Geng, Jiewen Ran, Xingguang Zhou, Chen Xing,
  Pengfei Wan, and Xiangyang Ji.
\newblock Range scaling global u-net for perceptual image enhancement on mobile
  devices.
\newblock In {\em Proceedings of the European Conference on Computer Vision
  (ECCV) Workshops}, pages 0--0, 2018.

\bibitem{hui2018perception}
Zheng Hui, Xiumei Wang, Lirui Deng, and Xinbo Gao.
\newblock Perception-preserving convolutional networks for image enhancement on
  smartphones.
\newblock In {\em Proceedings of the European Conference on Computer Vision
  (ECCV) Workshops}, pages 0--0, 2018.

\bibitem{ignatov2021fastDenoising}
Andrey Ignatov, Kim Byeoung-su, and Radu Timofte.
\newblock Fast camera image denoising on mobile gpus with deep learning, mobile
  ai 2021 challenge: Report.
\newblock In {\em Proceedings of the IEEE/CVF Conference on Computer Vision and
  Pattern Recognition Workshops}, pages 0--0, 2021.

\bibitem{ignatov2017dslr}
Andrey Ignatov, Nikolay Kobyshev, Radu Timofte, Kenneth Vanhoey, and Luc
  Van~Gool.
\newblock Dslr-quality photos on mobile devices with deep convolutional
  networks.
\newblock In {\em Proceedings of the IEEE International Conference on Computer
  Vision}, pages 3277--3285, 2017.

\bibitem{ignatov2018wespe}
Andrey Ignatov, Nikolay Kobyshev, Radu Timofte, Kenneth Vanhoey, and Luc
  Van~Gool.
\newblock Wespe: weakly supervised photo enhancer for digital cameras.
\newblock In {\em Proceedings of the IEEE Conference on Computer Vision and
  Pattern Recognition Workshops}, pages 691--700, 2018.

\bibitem{ignatov2021fastDepth}
Andrey Ignatov, Grigory Malivenko, David Plowman, Samarth Shukla, and Radu
  Timofte.
\newblock Fast and accurate single-image depth estimation on mobile devices,
  mobile ai 2021 challenge: Report.
\newblock In {\em Proceedings of the IEEE/CVF Conference on Computer Vision and
  Pattern Recognition Workshops}, pages 0--0, 2021.

\bibitem{ignatov2021fastSceneDetection}
Andrey Ignatov, Grigory Malivenko, and Radu Timofte.
\newblock Fast and accurate quantized camera scene detection on smartphones,
  mobile ai 2021 challenge: Report.
\newblock In {\em Proceedings of the IEEE/CVF Conference on Computer Vision and
  Pattern Recognition Workshops}, pages 0--0, 2021.

\bibitem{ignatov2020rendering}
Andrey Ignatov, Jagruti Patel, and Radu Timofte.
\newblock Rendering natural camera bokeh effect with deep learning.
\newblock In {\em Proceedings of the IEEE/CVF Conference on Computer Vision and
  Pattern Recognition Workshops}, pages 418--419, 2020.

\bibitem{ignatov2019aimBokeh}
Andrey Ignatov, Jagruti Patel, Radu Timofte, Bolun Zheng, Xin Ye, Li Huang,
  Xiang Tian, Saikat Dutta, Kuldeep Purohit, Praveen Kandula, et~al.
\newblock Aim 2019 challenge on bokeh effect synthesis: Methods and results.
\newblock In {\em 2019 IEEE/CVF International Conference on Computer Vision
  Workshop (ICCVW)}, pages 3591--3598. IEEE, 2019.

\bibitem{ignatov2019ntire}
Andrey Ignatov and Radu Timofte.
\newblock Ntire 2019 challenge on image enhancement: Methods and results.
\newblock In {\em Proceedings of the IEEE/CVF Conference on Computer Vision and
  Pattern Recognition Workshops}, pages 0--0, 2019.

\bibitem{ignatov2018ai}
Andrey Ignatov, Radu Timofte, William Chou, Ke Wang, Max Wu, Tim Hartley, and
  Luc Van~Gool.
\newblock Ai benchmark: Running deep neural networks on android smartphones.
\newblock In {\em Proceedings of the European Conference on Computer Vision
  (ECCV) Workshops}, pages 0--0, 2018.

\bibitem{ignatov2021real}
Andrey Ignatov, Radu Timofte, Maurizio Denna, and Abdel Younes.
\newblock Real-time quantized image super-resolution on mobile npus, mobile ai
  2021 challenge: Report.
\newblock In {\em Proceedings of the IEEE/CVF Conference on Computer Vision and
  Pattern Recognition Workshops}, pages 0--0, 2021.

\bibitem{ignatov2019aim}
Andrey Ignatov, Radu Timofte, Sung-Jea Ko, Seung-Wook Kim, Kwang-Hyun Uhm,
  Seo-Won Ji, Sung-Jin Cho, Jun-Pyo Hong, Kangfu Mei, Juncheng Li, et~al.
\newblock Aim 2019 challenge on raw to rgb mapping: Methods and results.
\newblock In {\em 2019 IEEE/CVF International Conference on Computer Vision
  Workshop (ICCVW)}, pages 3584--3590. IEEE, 2019.

\bibitem{ignatov2019ai}
Andrey Ignatov, Radu Timofte, Andrei Kulik, Seungsoo Yang, Ke Wang, Felix Baum,
  Max Wu, Lirong Xu, and Luc Van~Gool.
\newblock Ai benchmark: All about deep learning on smartphones in 2019.
\newblock In {\em 2019 IEEE/CVF International Conference on Computer Vision
  Workshop (ICCVW)}, pages 3617--3635. IEEE, 2019.

\bibitem{ignatov2020aimBokeh}
Andrey Ignatov, Radu Timofte, Ming Qian, Congyu Qiao, Jiamin Lin, Zhenyu Guo,
  Chenghua Li, Cong Leng, Jian Cheng, Juewen Peng, et~al.
\newblock Aim 2020 challenge on rendering realistic bokeh.
\newblock In {\em European Conference on Computer Vision}, pages 213--228.
  Springer, 2020.

\bibitem{ignatov2018pirm}
Andrey Ignatov, Radu Timofte, Thang Van~Vu, Tung Minh~Luu, Trung X~Pham, Cao
  Van~Nguyen, Yongwoo Kim, Jae-Seok Choi, Munchurl Kim, Jie Huang, et~al.
\newblock Pirm challenge on perceptual image enhancement on smartphones:
  Report.
\newblock In {\em Proceedings of the European Conference on Computer Vision
  (ECCV) Workshops}, pages 0--0, 2018.

\bibitem{ignatov2020aim}
Andrey Ignatov, Radu Timofte, Zhilu Zhang, Ming Liu, Haolin Wang, Wangmeng Zuo,
  Jiawei Zhang, Ruimao Zhang, Zhanglin Peng, Sijie Ren, et~al.
\newblock Aim 2020 challenge on learned image signal processing pipeline.
\newblock {\em arXiv preprint arXiv:2011.04994}, 2020.

\bibitem{ignatov2020replacing}
Andrey Ignatov, Luc Van~Gool, and Radu Timofte.
\newblock Replacing mobile camera isp with a single deep learning model.
\newblock In {\em Proceedings of the IEEE/CVF Conference on Computer Vision and
  Pattern Recognition Workshops}, pages 536--537, 2020.

\bibitem{ignatov2020controlling}
Dmitry Ignatov and Andrey Ignatov.
\newblock Controlling information capacity of binary neural network.
\newblock {\em Pattern Recognition Letters}, 138:276--281, 2020.

\bibitem{jacob2018quantization}
Benoit Jacob, Skirmantas Kligys, Bo Chen, Menglong Zhu, Matthew Tang, Andrew
  Howard, Hartwig Adam, and Dmitry Kalenichenko.
\newblock Quantization and training of neural networks for efficient
  integer-arithmetic-only inference.
\newblock In {\em Proceedings of the IEEE Conference on Computer Vision and
  Pattern Recognition}, pages 2704--2713, 2018.

\bibitem{jain2019trained}
Sambhav~R Jain, Albert Gural, Michael Wu, and Chris~H Dick.
\newblock Trained quantization thresholds for accurate and efficient
  fixed-point inference of deep neural networks.
\newblock {\em arXiv preprint arXiv:1903.08066}, 2019.

\bibitem{kim2020pynet}
Byung-Hoon Kim, Joonyoung Song, Jong~Chul Ye, and JaeHyun Baek.
\newblock Pynet-ca: enhanced pynet with channel attention for end-to-end mobile
  image signal processing.
\newblock In {\em European Conference on Computer Vision}, pages 202--212.
  Springer, 2020.

\bibitem{lee2018techology}
Yen-Lin Lee, Pei-Kuei Tsung, and Max Wu.
\newblock Techology trend of edge ai.
\newblock In {\em 2018 International Symposium on VLSI Design, Automation and
  Test (VLSI-DAT)}, pages 1--2. IEEE, 2018.

\bibitem{li2019learning}
Yawei Li, Shuhang Gu, Luc~Van Gool, and Radu Timofte.
\newblock Learning filter basis for convolutional neural network compression.
\newblock In {\em Proceedings of the IEEE/CVF International Conference on
  Computer Vision}, pages 5623--5632, 2019.

\bibitem{liu2018deep}
Hanwen Liu, Pablo Navarrete~Michelini, and Dan Zhu.
\newblock Deep networks for image-to-image translation with mux and demux
  layers.
\newblock In {\em Proceedings of the European Conference on Computer Vision
  (ECCV) Workshops}, pages 0--0, 2018.

\bibitem{liu2019metapruning}
Zechun Liu, Haoyuan Mu, Xiangyu Zhang, Zichao Guo, Xin Yang, Kwang-Ting Cheng,
  and Jian Sun.
\newblock Metapruning: Meta learning for automatic neural network channel
  pruning.
\newblock In {\em Proceedings of the IEEE/CVF International Conference on
  Computer Vision}, pages 3296--3305, 2019.

\bibitem{liu2018bi}
Zechun Liu, Baoyuan Wu, Wenhan Luo, Xin Yang, Wei Liu, and Kwang-Ting Cheng.
\newblock Bi-real net: Enhancing the performance of 1-bit cnns with improved
  representational capability and advanced training algorithm.
\newblock In {\em Proceedings of the European conference on computer vision
  (ECCV)}, pages 722--737, 2018.

\bibitem{lugmayr2019unsupervised}
Andreas Lugmayr, Martin Danelljan, and Radu Timofte.
\newblock Unsupervised learning for real-world super-resolution.
\newblock In {\em 2019 IEEE/CVF International Conference on Computer Vision
  Workshop (ICCVW)}, pages 3408--3416. IEEE, 2019.

\bibitem{lugmayr2020ntire}
Andreas Lugmayr, Martin Danelljan, and Radu Timofte.
\newblock Ntire 2020 challenge on real-world image super-resolution: Methods
  and results.
\newblock In {\em Proceedings of the IEEE/CVF Conference on Computer Vision and
  Pattern Recognition Workshops}, pages 494--495, 2020.

\bibitem{obukhov2020t}
Anton Obukhov, Maxim Rakhuba, Stamatios Georgoulis, Menelaos Kanakis, Dengxin
  Dai, and Luc Van~Gool.
\newblock T-basis: a compact representation for neural networks.
\newblock In {\em International Conference on Machine Learning}, pages
  7392--7404. PMLR, 2020.

\bibitem{romero2021real}
Andres Romero, Andrey Ignatov, Heewon Kim, and Radu Timofte.
\newblock Real-time video super-resolution on smartphones with deep learning,
  mobile ai 2021 challenge: Report.
\newblock In {\em Proceedings of the IEEE/CVF Conference on Computer Vision and
  Pattern Recognition Workshops}, pages 0--0, 2021.

\bibitem{ronneberger2015u}
Olaf Ronneberger, Philipp Fischer, and Thomas Brox.
\newblock U-net: Convolutional networks for biomedical image segmentation.
\newblock In {\em International Conference on Medical image computing and
  computer-assisted intervention}, pages 234--241. Springer, 2015.

\bibitem{silva2020deep}
Jose Ivson~S Silva, Gabriel~G Carvalho, Marcel~Santana Santos, Diego~JC
  Santiago, Lucas~Pontes de Albuquerque, Jorge F~Puig Battle, Gabriel~M da
  Costa, and Tsang~Ing Ren.
\newblock A deep learning approach to mobile camera image signal processing.
\newblock In {\em Anais Estendidos do XXXIII Conference on Graphics, Patterns
  and Images}, pages 225--231. SBC, 2020.

\bibitem{tan2019mnasnet}
Mingxing Tan, Bo Chen, Ruoming Pang, Vijay Vasudevan, Mark Sandler, Andrew
  Howard, and Quoc~V Le.
\newblock Mnasnet: Platform-aware neural architecture search for mobile.
\newblock In {\em Proceedings of the IEEE/CVF Conference on Computer Vision and
  Pattern Recognition}, pages 2820--2828, 2019.

\bibitem{TensorFlowLite2021}
TensorFlow-Lite.
\newblock https://www.tensorflow.org/lite.

\bibitem{timofte2018ntire}
Radu Timofte, Shuhang Gu, Jiqing Wu, and Luc Van~Gool.
\newblock Ntire 2018 challenge on single image super-resolution: Methods and
  results.
\newblock In {\em Proceedings of the IEEE conference on computer vision and
  pattern recognition workshops}, pages 852--863, 2018.

\bibitem{truong2021learning}
Prune Truong, Martin Danelljan, Luc Van~Gool, and Radu Timofte.
\newblock Learning accurate dense correspondences and when to trust them.
\newblock {\em arXiv preprint arXiv:2101.01710}, 2021.

\bibitem{uhlich2019mixed}
Stefan Uhlich, Lukas Mauch, Fabien Cardinaux, Kazuki Yoshiyama, Javier~Alonso
  Garcia, Stephen Tiedemann, Thomas Kemp, and Akira Nakamura.
\newblock Mixed precision dnns: All you need is a good parametrization.
\newblock {\em arXiv preprint arXiv:1905.11452}, 2019.

\bibitem{vu2018fast}
Thang Vu, Cao Van~Nguyen, Trung~X Pham, Tung~M Luu, and Chang~D Yoo.
\newblock Fast and efficient image quality enhancement via desubpixel
  convolutional neural networks.
\newblock In {\em Proceedings of the European Conference on Computer Vision
  (ECCV) Workshops}, pages 0--0, 2018.

\bibitem{wan2020fbnetv2}
Alvin Wan, Xiaoliang Dai, Peizhao Zhang, Zijian He, Yuandong Tian, Saining Xie,
  Bichen Wu, Matthew Yu, Tao Xu, Kan Chen, et~al.
\newblock Fbnetv2: Differentiable neural architecture search for spatial and
  channel dimensions.
\newblock In {\em Proceedings of the IEEE/CVF Conference on Computer Vision and
  Pattern Recognition}, pages 12965--12974, 2020.

\bibitem{wang2018esrgan}
Xintao Wang, Ke Yu, Shixiang Wu, Jinjin Gu, Yihao Liu, Chao Dong, Yu Qiao, and
  Chen Change~Loy.
\newblock Esrgan: Enhanced super-resolution generative adversarial networks.
\newblock In {\em Proceedings of the European Conference on Computer Vision
  (ECCV) Workshops}, pages 0--0, 2018.

\bibitem{wu2019fbnet}
Bichen Wu, Xiaoliang Dai, Peizhao Zhang, Yanghan Wang, Fei Sun, Yiming Wu,
  Yuandong Tian, Peter Vajda, Yangqing Jia, and Kurt Keutzer.
\newblock Fbnet: Hardware-aware efficient convnet design via differentiable
  neural architecture search.
\newblock In {\em Proceedings of the IEEE/CVF Conference on Computer Vision and
  Pattern Recognition}, pages 10734--10742, 2019.

\bibitem{yang2019quantization}
Jiwei Yang, Xu Shen, Jun Xing, Xinmei Tian, Houqiang Li, Bing Deng, Jianqiang
  Huang, and Xian-sheng Hua.
\newblock Quantization networks.
\newblock In {\em Proceedings of the IEEE/CVF Conference on Computer Vision and
  Pattern Recognition}, pages 7308--7316, 2019.

\end{thebibliography}

}

\end{document}